# Units of genetic transfer in prokaryotes

Cheong Xin Chan[1], Robert G. Beiko[2] and Mark A. Ragan[1]

[1]ARC Centre of Excellence in Bioinformatics, The University of Queensland, Brisbane QLD 4072, Australia

[2]Department of Computer Science, Dalhousie University, 6050 University Avenue, Halifax, Nova Scotia, Canada B3H 1W5

**Abstract**

The transfer of genetic materials across species (lateral genetic transfer, LGT) contributes to genomic and physiological innovation in prokaryotes. The extent of LGT in prokaryotes has been examined in a number of studies, but the unit of transfer has not been studied in a rigorous manner. Using a rigorous phylogenetic approach, we analysed the units of LGT within families of single-copy genes obtained from 144 fully sequenced prokaryote genomes. A total of 30.3% of these gene families show evidence of LGT. We found that the transfer of gene fragments has been more frequent than the transfer of entire genes, suggesting the extent of LGT has been underestimated. We found little functional bias between within-gene (fragmentary) and whole-gene (non-fragmentary) genetic transfer, but non-fragmentary transfer has been more frequent into pathogens than into non-pathogens. As gene families that contain probable paralogs were excluded from the current study, our results may still underestimate the extent of LGT; nonetheless this is the most-comprehensive study to date of the unit of LGT among prokaryote genomes.

# Introduction

In prokaryotes, exchange of genetic material between lineages can counteract the accumulation of deleterious mutations, replacing damaged DNA and helping to maintain genetic variation. In some cases the introgressed genetic material may confer a selective advantage to the new host organism, resulting in positive selection. One well-known example of this is the acquisition and spread of genes encoding antibiotic resistance in highly selective environments [1]. In the process, though, phylogenetic histories become entangled, and the very concept of a genomic or species phylogeny becomes fraught [2-4].

The inheritance of genetic materials in prokaryotes is largely vertical, *i.e.* transmitted from parent to offspring within a genomic and organismal lineage. However, a number of large-scale studies have identified substantial evidence for LGT. Some of these are based on the topological comparison of phylogenetic trees inferred for individual gene families, *e.g.* against a reference topology [5, 6]. The most careful such study published so far showed that organisms that are closely related phylogenetically, and/or are found in a common environmental niche, show a tendency to share genetic material *via* LGT [5]. Other approaches to quantify the extent of LGT include the examination of nucleotide composition or codon usage patterns [7], inference of gene gain and loss events [8, 9], and calculations of ancestral genome sizes under the assumption that in the absence of LGT, present-day diversity must be shared and derived from the common ancestral genome [10].

Because it is difficult to detect introgressed genomic regions that originate from closely related lineages (those with a high degree of sequence identity), the regions most confidently inferred to be of lateral origin may often be those that have come from more distantly related sources, perhaps *via* transduction through phage [11, 12]. However, sequences can be divergent not only due to temporal separation from their common ancestor, but also because they have become functionally specialised, as is often the case with paralogs. Following duplication, a genetic region can lose its original function (non-functionalisation), gain a novel function (neofunctionalisation), or take on a specialised part of the original function (subfunctionalisation) [13]. As these processes can of course take place not only in the new host lineage but also in candidate donor lineages, paralogy can complicate the inference of lateral transfer (and *vice-versa*).

In the process of LGT, exogenous genetic materials are first introduced into the recipient cell, and then integrated into the new host *via* recombination. The integrated genetic material can constitute an entire gene [14], a partial (fragmentary) gene [15, 16], or multiple (entire or fragmentary) adjacent genes [17, 18]. Although several studies have explored the frequency of LGT in prokaryotes and examined individual genes or functions affected [7, 19], none of these has taken a comprehensive rigorous approach to characterising the *unit*s of genetic transfer. Given the large number of completely sequenced prokaryote genomes now available in the public domain, such an analysis is now possible and timely.

Here we report results of the first systematic study of the unit of lateral genetic transfer across the diversity of sequenced prokaryote genomes. We characterise the frequencies of within- and entire-gene transfer, and discuss correlations with annotated gene functions and phyletic group. To minimise, to the extent possible, the complications of paralogy and to increase the confidence with which we can infer LGT events, we focus here on families of single-copy genes.

# Results

For discovery of LGT events in prokaryote genomes we extracted a subset of the 22437 putatively orthologous families used in our previous large-scale study [5] on LGT in 144 phyletically diverse prokaryote genomes (see Materials and Methods). This subset, 1462 gene families, was restricted to families of single-copy genes, *i.e.* genes that are sufficiently unique within their respective genomes to make it unlikely that they have arisen by gene duplication. By applying this restriction, we ensure to the extent possible that any recombination we infer in any of these families arises from LGT, not paralogy. The size distribution of these 1462 gene families is shown in Figure 1.

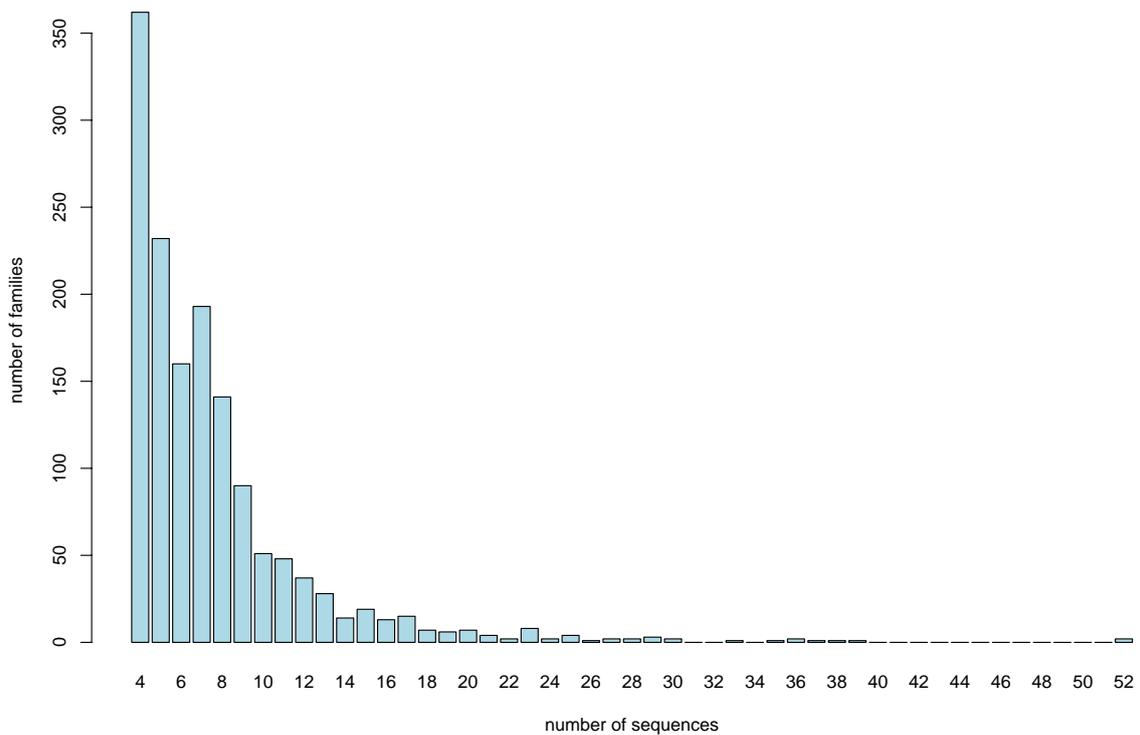

Figure 1. Size distribution of families of single-copy genes examined in this study.

Family sizes range from 4 to 52 members; 1229 (84.1%) of the families contain ≤ 10 sequences, with almost a quarter of them (362, 24.7%) of size 4. Gene families of size < 4 were excluded from the analysis, as they do not contribute to meaningful phylogenetic inference. Each of the 1462 families was examined for evidence of LGT, as described below.

*Within-gene (fragmentary) genetic transfer*

We applied a two-phase strategy [20] for detecting recombination in each of the 1462 families of single-copy genes. In the first phase we used three statistical measures [21-23] to search for evidence of phylogenetic discrepancies (*i.e.* a recombination signal) within the family; recombination was inferred if two of the three tests show a *p*-value ≤ 0.10. In the second phase we utilised a Bayesian phylogenetic approach, implemented in the software program DualBrothers [24], to locate recombination breakpoints more precisely in the families that, in the first phase, showed evidence of recombination. DualBrothers employs reversible-jump Markov chain Monte Carlo (MCMC) and a dual multiple change-point model to identify, within a set of sequences, contiguous regions that share a common tree topology, and the boundaries (recombination breakpoints) between regions that show different topologies [24, 25].

Instances of recombination discovered using this approach are thus necessarily fragmentary, as at least one end of a topologically distinct region (*i.e.* a recombination breakpoint) occurs within the sequence set used in our analysis. Whole-gene transfer escapes detection because it does not result in topological discrepancy along the length of these sequences.

Our first-phase screening produced evidence of recombination in 426 (29.1%) of these 1462 families. Of these, we found clear evidence of recombination in 286 (19.6%), where "clear evidence" is defined as Bayesian posterior probability (BPP) support ≥ 0.500 for the dominant topology on at least one side of the inferred breakpoint. We found a further 80 cases (5.5%) in which a breakpoint was located, but no sequence region has BPP ≥ 0.500; we classified these as inconclusive. Finally, we observed 60 cases (4.1%) for which recombination was indicated in the initial screening, but no recombination breakpoint could be identified. First-phase screening did not detect recombination in 1036 families (70.9%).

Figure 2 shows the size distribution of these 286 gene families; the most-populated classes are of eight (42 families, 14.7%) and six sequences each (39 families, 13.6%).

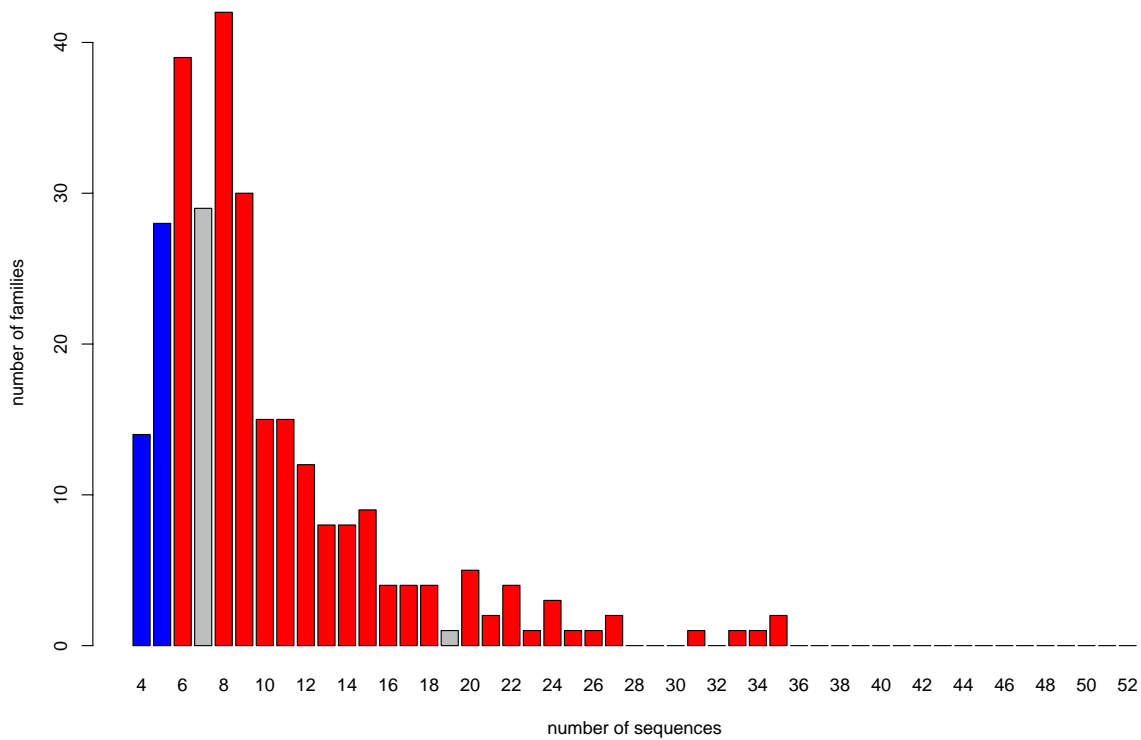

Figure 2. Size distribution of gene families that show evidence of fragmentary lateral genetic transfer. The red bars indicate over-represented (frequency more-than-expected) groups; the blue bars indicate under-represented (frequency less-than-expected) groups; and the grey bars indicate groups indifferent (neither over- nor under-represented) in comparison with the 1462-family dataset, at $p \leq 0.05$.

Among these 286 gene families, those with ≤ 5 members are under-represented ($p \leq 0.05$) based on their frequencies in the 1462-family dataset, implying that fragmentary LGT in these smallest families either (a) has been less frequent than that into the larger families, or (b) is more difficult to detect. Conversely, almost all gene families of size > 5 are individually over-represented.

*Whole-gene (non-fragmentary) genetic transfer*

We next inferred phylogenetic trees for each of the 1096 gene families for which no recombination inferred (1036 from the first phase, 60 from the second), and compared the inferred topology with a reference tree. The reference (species) tree [5] was generated using Matrix Representation with Parsimony (MRP) [26], yielding a supertree that summarises all well-supported (BPP ≥ 0.95) bipartitions among the 22432 trees of putatively orthologous families in these 144 prokaryote genomes [5]. In the absence of a detectable recombination breakpoint within the gene, phylogenetic discordance between a well-supported gene tree and the reference supertree can most readily be interpreted as lateral transfer the entire gene (and probably beyond).

We found 157 gene families of single-copy genes (10.7% of the 1462-family dataset) that are topologically incongruent with the reference tree, suggesting that non-fragmentary genetic transfer had affected these families. Their size distribution is depicted in Figure 3. As in the fragmentary transfer cases above (Figure 2), small gene families (size ≤ 5) are under-represented relative to the others, while two-thirds of the families of size ≥ 8 are over-represented ($p \leq 0.05$).

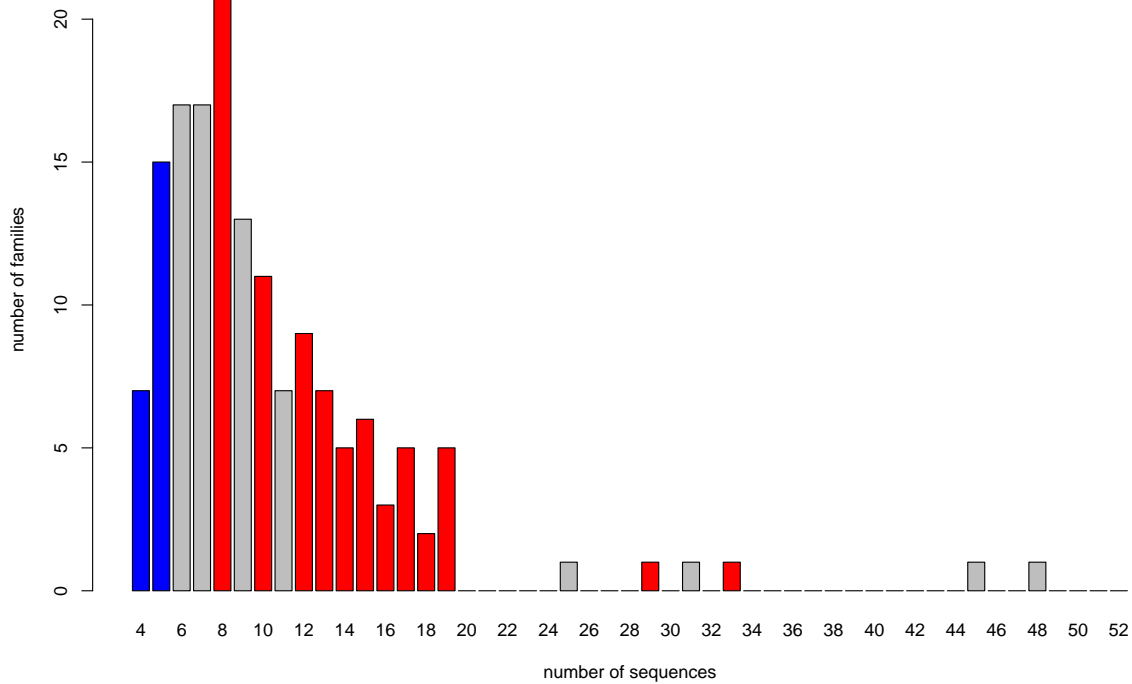

Figure 3. Size distribution of gene families that show evidence of non-fragmentary lateral genetic transfer. The red bars indicate over-represented (more-than-expected) groups; the blue bars indicate under-represented (less-than-expected) groups; and the grey bars indicate groups indifferent (neither over- nor under-represented) in comparison with the 1462-family dataset, at $p \leq 0.05$.

In total, among the 1462 families of single-copy genes we found evidence of LGT in 443 (30.3%), of which 286 (64.5%) show within-gene (fragmentary) and 157 (35.4%) whole-gene (non-fragmentary) recombination.

*Functional biases of fragmentary and non-fragmentary genetic transfer*

We used annotations from the TIGR Comprehensive Microbial Resource (http://cmr.tigr.org/) to assign a functional category (TIGR role category) to the protein associated with each gene in these 443 gene families. Details are provided in Materials and Methods. Figure 4 shows the proportions of proteins in each functional category, broken out by membership in families for which we inferred (a) fragmentary and (b) non-fragmentary LGT.

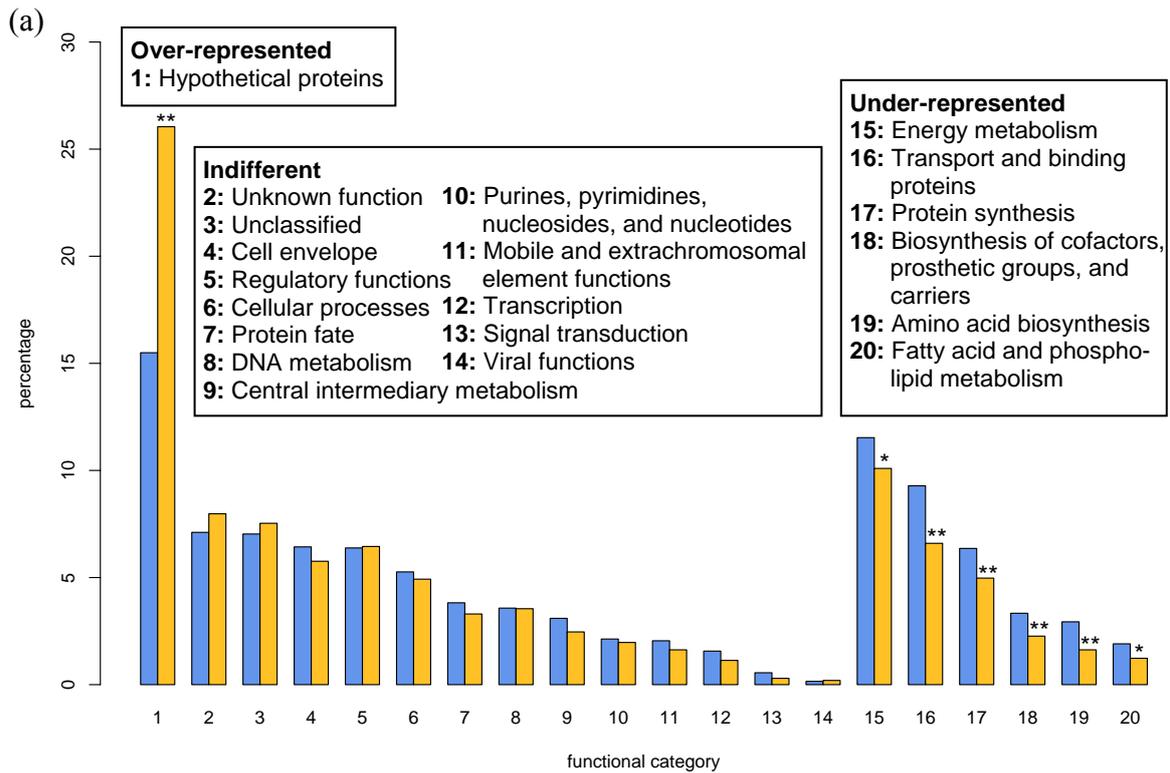

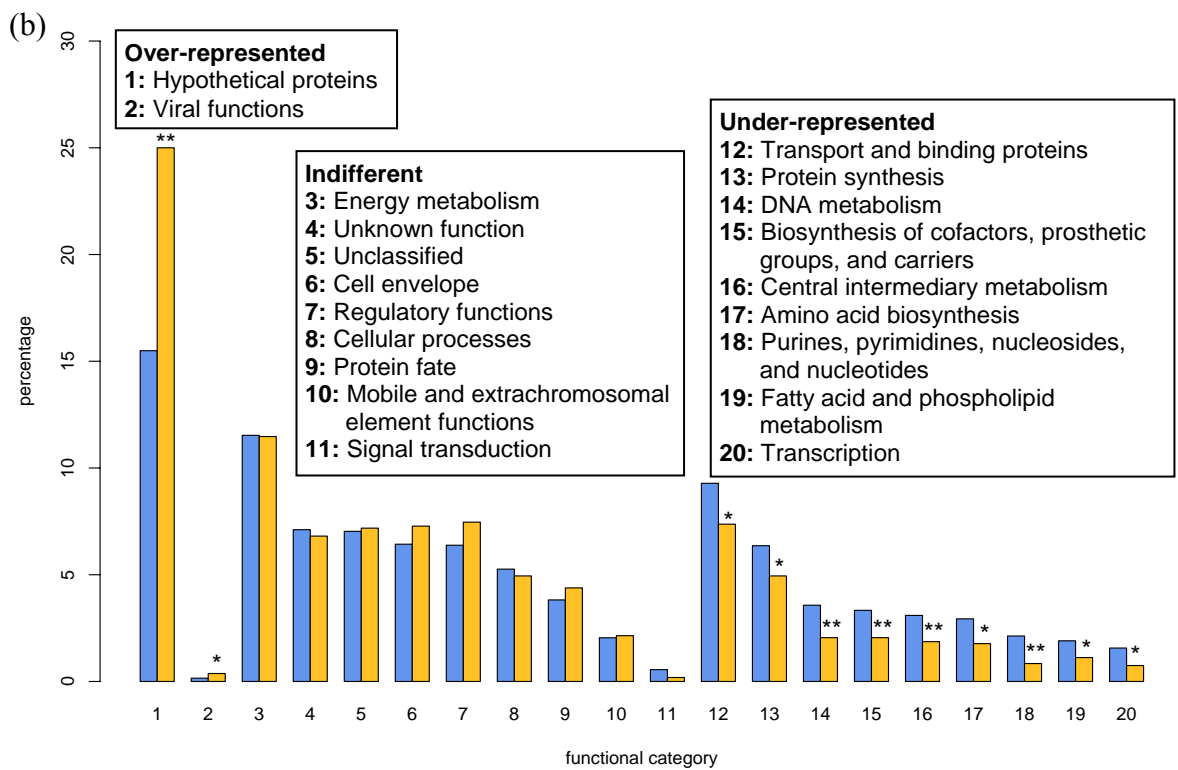

Figure 4. Representation of functional categories assigned to protein sequences corresponding to gene families that show evidence of (a) fragmentary and (b) non-fragmentary genetic transfer (yellow bars). The blue bars show the representation of these same functional categories in the full dataset (16639 families of size $\geq 4$, 119695 proteins). Categories are numbered (differently for panels a and b) as shown in the boxes. Significance of over- or under-representation is represented by single ($p \leq 0.05$) and double asterisks ($p \leq 0.01$).

Hypothetical proteins constitute the major over-represented category, both as a proportion of proteins corresponding to families for which we infer within-gene recombination, and as a proportion of proteins corresponding to families in which one or more entire genes has arisen by LGT. A relatively tiny category of proteins related to viral functions (including transduction of DNA by phages) is the only other category similarly over-represented, and it is over-represented only among proteins corresponding to families for which we infer non-fragmentary transfer. On the contrary, proteins involved in a range of biosynthetic, metabolic, protein-synthetic, transport and binding functions are significantly under-represented in both within-gene and whole-gene transfer. Proteins that function in energy metabolism are under-represented only in the case of fragmentary transfer (Figure 4a), while those engaged in DNA metabolism, central intermediary metabolism, and transcription are under-represented only for non-fragmentary LGT (Figure 4b).

***Phyletic biases of fragmentary and non-fragmentary genetic transfer***

We next asked whether within-gene and whole-gene lateral transfer is over- or under-represented in particular taxa. Figure 5 shows the taxonomic origins (NCBI level-4 taxa) of proteins that correspond to families within which we infer fragmentary and non-fragmentary genetic transfer. For clarity, the corresponding proportions are not shown over the entire 144-genome (16639-family) dataset; over- and under-representation ($p \leq 0.05$) are indicated by red and blue colouration respectively.

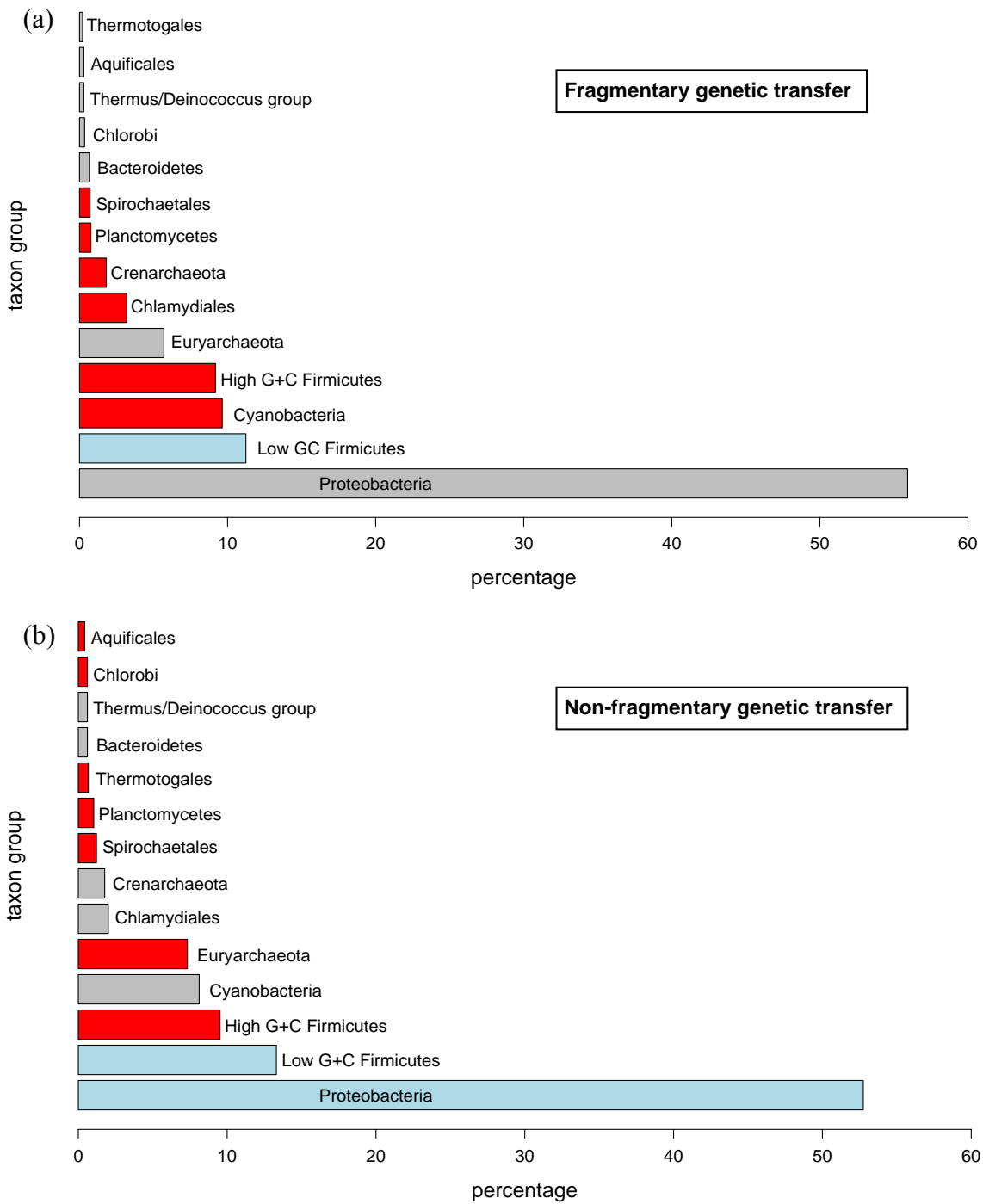

Figure 5. Taxonomic origins (NCBI level-4 taxa) of genes in families that show evidence of (a) fragmentary and (b) non-fragmentary genetic transfer. Over-representation relative to the 16639-family dataset is shown in red; under-representation is shown in blue; grey indicates that there is neither over- nor under-representation at $p \leq 0.05$.

Our results reveal that families of single-copy genes affected by LGT contain a significantly ($p \leq 0.05$) higher-than-expected proportion of genes originating from High-G+C Firmicutes, Planctomycetes and Spirochaetales. This is true for both the fragmentary and non-fragmentary transfer cases.

Other taxonomic groups are over-represented in only the fragmentary transfer case, or only the non-fragmentary, but not both. The data and our approach do not allow us to extrapolate with certainty, but to the extent that these single-gene families are representative of complete genomes, the cyanobacteria, chlamydiales and crenarchaeotes appear to be relatively receptive to introgression of gene fragments, whereas euryarchaeotes, chlorobi, and members of *Thermotoga* and *Aquifex* have been relatively receptive to transfer of entire genes. We note that many of the latter taxa are extremophiles, suggesting that it may bear further analysis whether whole-gene transfer is more frequent than fragmentary genetic transfer among organisms that live in *e.g.* high-temperature and highly saline environments.

Low-G+C Firmicutes are under-represented in families affected by both types of LGT, fragmentary and non-fragmentary. On the other hand Proteobacteria, the high-level taxon most abundantly represented in our dataset, is significantly under-represented in families affected by whole-gene transfer.

Table 1 shows the species that are significantly over-represented in families within which we infer recombination in this dataset. Many over-represented species are pathogens.

Table 1. Species that are over-represented ($p \leq 0.05$) in gene families that show evidence of genetic transfer, in comparison with their contribution to the 16639-family dataset. Species are listed in descending order, from the most over-represented to the least over-represented, separately for fragmentary and non-fragmentary genetic transfer. The species represented in red are pathogens. The four species listed separately at the bottom of the Table are over-represented in both fragmentary and non-fragmentary cases.

| **Fragmentary genetic transfer** | **Non-fragmentary genetic transfer** |
|---|---|
| *Nostoc sp.* PCC 7120 | *Salmonella typhimurium* LT2 |
| *Streptomyces avermitilis* MA-4680 | *Salmonella enterica* subsp. *enterica* serovar Typhi Ty2 |
| *Shewanella oneidensis* MR-1 | *Mycobacterium tuberculosis* CDC1551 |
| *Yersinia pestis* CO92 | *Nitrosomonas europaea* ATCC 19718 |
| *Synechococcus sp.* WH 8102 | *Yersinia pestis* KIM |
| *Thermosynechococcus elongatus* BP-1 | *Methanothermobacter thermautotrophicus* |
| *Pasteurella multocida* | *Leptospira interrogans* serovar lai str. 56601 |
| *Methanococcus jannaschii* | *Thermotoga maritima* |
| *Methanopyrus kandleri* AV19 | *Fusobacterium nucleatum* subsp. *nucleatum* ATCC 25586 |
| *Halobacterium sp.* NRC-1 | *Chlorobium tepidum* TLS |
| *Chlamydophila pneumoniae* CWL029 | *Coxiella burnetii* RSA 493 |
| *Mycoplasma pulmonis* | *Aquifex aeolicus* |
|  | *Treponema pallidum* |
|  | *Streptococcus pyogenes* MGAS315 |
| *Photorhabdus luminescens* subsp. *laumondii* TTO1 | |
| *Haemophilus ducreyi* 35000HP | |
| *Pirellula* sp. | |
| *Borrelia burgdorferi* | |

## Discussion

Our results demonstrate that in these families of single-copy genes from diverse prokaryotes, transfer of genetic material is largely vertical, but a significant proportion of gene families (30.3%) show clear evidence of LGT. In previous studies, estimates of the frequency of LGT range widely: 2% [27], 13% [5], 16% [8], 60% [6], to as high as 90% [28] of genes or bipartitions. In a recent study based on inference of ancestral genome sizes [29], all genes in prokaryotes were proposed to have had undergone LGT. Several factors contribute to this range of estimates, including but not limited to methodological approach and sampling of genes and genomes. Different methodologies can produce not only different estimates of the extent of LGT, but incompatible lists of lateral genes, on the same dataset [30]. The phylogenetic approach to detection of LGT is firmly grounded in biological principle (the same principles as those responsible for inheritance and diversification of lineages) and can be carried out in a statistically rigorous manner, although systematic biases, *e.g.* surrounding the model of sequence change, may still intrude.

A limitation of the phylogenetic approach as adopted in previous studies [5, 6], however, has been the intrinsic assumption that the unit of genetic transfer is a whole gene. Topological discordance between a gene-family tree and the reference topology has been interpreted as *prima facie* evidence that a gene has been transferred from one lineage into another. Here, we have employed a phylogenetic approach but without restricting the unit of transfer to be a whole gene, and have shown that among these diverse prokaryotic species, LGT can involve the recombination of a fragment smaller than a gene and/or the interruption of an existing gene. Indeed, over the set of families of single-copy genes in these genomes, within-gene transfer is about twice as frequent as the transfer of entire genes (or larger).

The dataset used in this study is a subset of that used by Beiko *et al.* [5], who concluded that some 13-14% of bipartitions are affected by LGT. Here we report 30% of families are affected by LGT. These two numbers are not directly comparable, for three reasons: (1) our present subset is non-representative, comprising only families of single-copy genes; (2) our present dataset is smaller, having 74% as many families and 54% as many sequences; and (3) we base our analyses on gene families, not on bipartitions, as genetic transmission involving within-gene recombination can only partially be mapped into the paradigm of bipartitions and subtrees. Neither we nor Beiko *et al.* [5] attempted to estimate LGT in paralogous families, or among very closely related genomes. Again we are reminded of the multifaceted trade-offs between methodological rigor, and the goal of a more-global estimate of frequency of LGT in prokaryotes.

Extrapolation of results from families of single-copy genes to entire genomes might be on the least-solid ground in Proteobacteria, which in other studies have been found to show high rates of LGT [3, 31]. Gene duplications are more common among Proteobacteria than among many other prokaryotes [32]. However, as discussed elsewhere, in this study we excluded families with duplicates in individual genomes.

Here we have also shown that LGT is less evident in small gene families ($N \leq 5$) than in larger gene families. In our data, gene family size is correlated with degree of sequence divergence, as many small families represent closely related organisms (genomes) that have only recently diverged from a common ancestor [33]. As it is difficult to detect genetic transfer events involving highly similar sequences, the frequency of LGT among small gene families can be underestimated. Conversely, larger families typically are constituted by representatives from phyletically more-diverse organisms with a more-ancient common ancestor, making it is easier to detect phylogenetic discrepancies and hence to infer LGT [34].

We observed only modest differences in functional bias between fragmentary and non-fragmentary transfer in families of single-copy genes; hypothetical proteins are very significantly ($p \leq 0.01$) over-represented in both cases. Gene products not classified by the TIGR Comprehensive Microbial Resource, or of unknown function, are not significantly over- (or under-) represented relative to our full dataset, suggesting that an exogenous or hybrid origin does not significantly decrease (or increase) annotation of a functional role category. We also found that genes encoding viral functions are more likely to be laterally transferred in their entirety than as fragments. A similar trend is observed for pathogenic bacteria, which are prominent among the organisms that contribute disproportionately to families affected by non-fragmentary transfer. Genes that encode for virulence factors (*e.g.* toxins, adhesins and invasins) are known to be commonly located on mobile genetic elements such as plasmids and transposons, or in specific genomic region called pathogenicity islands [35, 36].

We observed that genes annotated as involved in DNA metabolism, transcription, and protein synthesis are under-represented among families for which we infer whole-gene LGT, although of these only the protein synthesis functional category is also under-represented in fragmentary transfer. The complexity hypothesis [37] postulates that "informational" proteins involved in processes related to transcription and translation, including many in these three categories, typically function in the cell within large multi-protein complexes and hence must interact in finely tuned ways with many other biomolecules, and as a consequence their genes are less likely to be susceptible to transfer *via* LGT than are genes encoding the putatively less-interactive "operational" proteins. However, the susceptibility of the genomes to transfer of "informational" genes can still be underestimated, especially among highly similar sequences, in which detection of recombination is difficult. Our results do not speak directly to the validity of this hypothesis, but suggest that any bias against transfer of informational genes may be expressed more strongly in the case of whole-gene than within-gene transfer.

## Materials and Methods

*Data*

From 144 completely sequenced prokaryote genomes we generated 22437 putatively orthologous protein families of size $N \geq 4$ *via* a hybrid clustering approach [38]. We aligned these families [5] and validated the alignments using a pattern-centric objective function [19]. These protein sequence alignments were converted into DNA sequence alignments by retrieving the corresponding nucleotide sequences from GenBank (http://www.ncbi.nlm.nih.gov/) and arranging the nucleotide triplets to parallel exactly the protein alignment in each case, yielding 18809 gene families ($N \geq 4$) containing a total of 139707 genes. We require $N \geq 4$ because 4 is the minimum size that can yield distinct topologies; however, this is true only if every sequence in the family has a unique sequence. Therefore we identified sets of identical sequences and removed (at random) all but one copy of each, yielding 16639 families ($N \geq 4$) and 119695 genes. In every case, the identical copies removed from consideration represented organisms either in the same genus (99.7%), or within the *Escherichia-Shigella* species pair (0.3%); many represent different strains within the same species (89.1%). It is possible that some of these represent (within-gene or whole-gene) LGT, but such cases could not have been detected by our (or any other existing) approach in any case.

To minimise, to the extent possible, erroneous inference arising from the presence of paralogous sequences within these families, we further restricted our dataset to those 1462 families for which each member represents a different genome. In this dataset, these families of single-copy genes range in size from 4 to 52 (Figure 1).

*Detecting fragmentary genetic transfer*

We applied a two-phase strategy for detecting recombination [20] in this study. In the first phase, PhiPack [21] was used to detect the occurrences of recombination based on discrepancies of phylogenetic signal within the sequence alignments. The program incorporates *p*-values of the NSS statistics in Reticulate [23], the MaxChi test [22], and PHI [21]. Datasets with at least two of the three *p*-values ≤ 0.10 were considered as positive for recombination.

In the second phase, for each sequence set that showed evidence of recombination (above), a Bayesian phylogenetic approach was used to delineate recombination breakpoints; this was implemented in DualBrothers [24] run with MCMC chain length = 2500000, burnin = 500000, window_length = 5, and Green's constant C = 0.25. The tree search space for each run of DualBrothers is defined by a list of unrooted tree topologies inferred using MRBAYES [39], for which we used parameter settings MCMC chain length = 2500000 and burnin = 500000 (nucmodel = 4by4, rates=gamma, ngammacat = 4) on smaller partitions of the sequence set, *via* a window-sliding approach (window length = 100, sliding size = 50, unit in alignment position); tree topologies within a 90% Bayesian confidence interval were included, with 1000 trees maximum. Gene families that show evidence of recombination are inferred to have undergone one or more events of fragmentary genetic transfer.

*Detecting non-fragmentary genetic transfer*

For each gene family for which no evidence of recombination was found in the first-phase screen, and for those positive in the first-phase screen but for which no recombination breakpoint could be detected, we inferred a Bayesian phylogenetic tree (see below) and compared its topology against that of a reference tree; whole-gene (non-fragmentary) genetic transfer was inferred if the topologies were significantly discordant. These individual gene-family trees were inferred from DNA alignments (above). As reference we used the MRP [26] computed from all well-supported (BPP ≥ 0.95) bipartitions among all individual protein-family trees in these 144 genomes [5]. The individual gene-family trees were inferred using MRBAYES [39] with MCMC chain length = 2500000, burnin = 500000, and model = K2P [40]. Possible discordance between individual gene-family trees and the reference supertree topology was assessed under likelihood models captured in the Shimodaira-Hasegawa test [41], the one- and two-sided Kishino-Hasegawa tests [42, 43], and expected likelihood weights [44], all as implemented in Tree-Puzzle 5.1 [45]. Discordance was inferred if any tree was rejected by more than two of the four ML tests at a confidence interval of 95% ($p \leq 0.05$), and was taken as *prima facie* evidence of whole-gene (non-fragmentary) lateral genetic transfer.

*Functional analysis of gene families*

Functional information for each protein sequence was retrieved from the Comprehensive Microbial Resource (CMR) at The Institute for Genomic Research website (http://cmr.tigr.org/), based on TIGR role identifiers and categorisation at Level 1. Over- or under-representation of functional categories and taxonomic groups was based on the probability of observing a defined number of target groups (or categories) in a subsample, given a process of sampling without replacement from the whole dataset (as defined in each case: see text) under a hypergeometric distribution [46]. The probability of observing $x$ number of a particular target category is described as:

$$P(k=x) = f(k;N,m,n) = \frac{\binom{m}{k}\binom{N-m}{n-k}}{\binom{N}{n}}$$

in which $N$ is the total population size, $m$ is the size of the target category within the population, $n$ is the total size of the subsample, and $k$ is the size of the target category within the subsample.

## Acknowledgements

This study was supported by Australian Research Council (ARC) grant CE0348221. We thank Aaron Darling and Vladimir Minin for valuable advice on the use of DualBrothers. CXC was supported by a University of Queensland UQIPRS scholarship.